\documentclass[12pt, nofootinbib,floatfix,amsmath,amssymb]{revtex4-2}
\usepackage{graphicx}
\usepackage{multirow}
\usepackage{amssymb}
\usepackage[usenames]{color}
\usepackage[toc,page]{appendix}
\usepackage{slashed}
\usepackage{bbding}
\usepackage{booktabs}
\usepackage{colordvi}
\usepackage{amsmath}
\usepackage{amssymb}
\usepackage{amsfonts}
\usepackage[dvipsnames]{xcolor}
\usepackage[colorlinks=true,urlcolor=blue,linkcolor=blue,citecolor=blue]{hyperref}
\usepackage{cancel}
\usepackage{caption}
\usepackage{subcaption}
\usepackage{float}
\usepackage{gensymb}
\usepackage{soul}
\usepackage[top=1.135in,bottom=1.17in,left=1.16in,right=1.16in]{geometry}
\numberwithin{equation}{section}

\allowdisplaybreaks

\newcommand{\be}{\begin{equation}}
\newcommand{\ee}{\end{equation}}

\newcommand{\eq}[1]{\begin{align}#1\end{align}}

\begin{document}


\title{$\tau^- \to \pi^- \eta \nu_\tau$ decay induced by QED one-loop effects}

\author{Gerardo Hernández-Tomé}%
 \email{gerardo.hernandez@cinvestav.mx}
\author{Gabriel López Castro}
\email{gabriel.lopez@cinvestav.mx} 
\author{Diego Portillo-Sánchez}
\email{diego.portillo@cinvestav.mx}
\affiliation{ {\it Departamento de F\'isica, Centro de Investigaci\'on y de Estudios Avanzados del Instituto Polit\'ecnico Nacional} \\ {\it Apartado Postal 14-740, 07000 Ciudad de M\'exico, M\'exico}
} %

\date{\today}

\begin{abstract}

The $\tau^- \to \pi^-\eta\nu_{\tau}$ decay is forbidden in the Standard Model in the limit of exact $G$-parity, it becomes a rare decay due to isospin symmetry breaking and it is very sensitive to the effects of effective scalar interactions. Since the parameters driving isospin breaking, $(m_d-m_u)/(m_s-\bar{m})$ and $\alpha$, are of the same order, one may expect their $G$-parity breaking effects in this decay can be of similar magnitudes. In this work, we evaluate the effects of isospin-breaking amplitudes originated from a virtual photon at one-loop in a resonance dominance model to describe photon-hadron interactions. We find that these effects can shift the leading SM predictions based on the $u-d$ quark mass difference by roughly 12\%, and should be taken into consideration in a precision comparison of theory and experiment in order to draw meaningful conclusions on New Physics. The effects in the rate of the analogous $\tau^- \to \pi^-\eta'\nu_{\tau}$ decay can be larger ($\sim$  78\%), under the approximations assumed in this model.
\end{abstract}

\maketitle
\newpage

\section{Introduction}

The study of rare and forbidden processes in the Standard Model (SM) is important because they can be sensitive to the effects of new particles or interactions. If rare decays are suppressed beyond experimental sensitivity, any positive signal would be due to New Physics (NP); in case they are at the reach of experimental searches, good control of SM prediction is necessary in order to extract meaningful information from the measured observables. The latter is the case of the rare $\tau^- \to \pi^-\eta\nu_{\tau}$ decay studied in this paper. As shown in Ref. \cite{Garces:2017jpz}, because this decay is forbidden by $G$-parity\footnote{$G\equiv Ce^{iI_3}$, where $C$ is the charge conjugation operator and $I_3$ the third component of the isospin operator  \cite{Lee:1956sw}. }, it can be very sensitive to the effects of dimension-six scalar interactions for low-energy semileptonic processes in the framework of an effective field theory.

In the SM of electroweak interactions, the strangeness-conserving semileptonic decays of $\tau^-$ leptons are mediated by the $(V-A)_{\mu}=\bar{d}\gamma_{\mu}(1-\gamma_5)u$ weak charged current. Owing to the $G$-parity properties of the vector (axial) current\footnote{The vector (axial) current of the $V-A$ theory was assigned a $G= +1 (-1)$ parity and were named by Weinberg \cite{Weinberg:1958ut} as `first class' currents, while the term `second class' was deserved to scalar (S) and pseudotensor (PT) currents with opposite $G$-parity. Although this terminology has become obsolete nowadays, in this paper we will refer sometimes to the non-standard S and PT interactions as second-class currents.} \cite{Weinberg:1958ut}, tau leptons can decay into final states that conserve $G$-parity, like an even (odd) number of pions. Therefore, in 1978 Leroy and Pestieau \cite{Leroy:1977pq} have suggested that the $\tau^-\to a_0^-(980)\nu_{\tau},\  b_1(1235)\nu_{\tau}$ decays, with the subsequent $a_0 \to \eta\pi^-, \  b_1\to \omega\pi^-$ would be good signals of non-SM currents since the $\eta\pi^-$ ($\omega\pi^-$) system has a $G$-parity quantum number opposite to that of the vector (axial) current.

The $\tau^-$ lepton decay of our concern has been calculated by many authors in the past four decades \cite{Leroy:1977pq, Tisserant:1982fc, Bramon:1987zb, Pich:1987qq, Diaz-Cruz:1991qbl, Bednyakov:1992af, Neufeld:1994eg, Nussinov:2008gx, Nussinov:2009sn, Paver:2010mz, Paver:2011md, Volkov:2012be, Descotes-Genon:2014tla, Escribano:2016ntp, Volkov:2021kus, Moussallam:2021flg}. The different predictions yield branching fractions in the range BR$(\tau^- \to \pi^-\eta\nu_\tau) \sim $ $O(10^{-5}\sim 10^{-6})$. The underlying mechanism in those model-dependent calculations is driven by the $m_d-m_u$ quark mass difference, either due to a first-class current followed by the $\pi^0-\eta$ mixing ($\tau^- \to \rho^-(\to \pi^-\pi^0\to \pi^-\eta) \nu_{\tau}$) or induced by isospin breaking (IB) in the weak vertex ($\tau^- \to a_0^-(\to \pi^-\eta )\nu_{\tau}$). Other calculations assume that $\tau^-\to \pi^-\eta\nu_{\tau}$ is mediated by NP in the form of scalar interactions \cite{Bramon:1987zb, Meurice:1987pp}. Given that $G$-parity violating effects make this a rare decay process, the contributions of NP may become competitive.

To the best of our knowledge, IB effects induced by electromagnetic interactions have been considered only in Ref. \cite{Hernandez-Tome:2017pdc}, which turned out to be a very small effect of $O(\alpha^2)$ at the amplitude level. In this paper, we consider the IB effects that are induced by electromagnetic interactions at the one-loop level, which leads to an amplitude suppressed only at $O(\alpha)$. Since the $\pi^0-\eta$ mixing parameter $\epsilon_{\eta\pi}$, as well as the fine structure constant $\alpha$ turn out to be of similar order (roughly $ 1\%$), one may expect {\it a priory} those effects may contribute to the amplitude at the same level.

Regarding the experimental searches for this rare tau decay, the first upper limits were reported in the nineties by the CLEO \cite{CLEO:1992nqz, CLEO:1996rmd} $\textrm{BR}(\tau^-\to\pi^-\eta\nu_\tau)<1.4\times 10^{-4}$ and ALEPH \cite{ALEPH:1996kok} $\textrm{BR}(\tau^-\to\pi^-\eta\nu_\tau)<6.2\times 10^{-4}$ collaborations. Those limits were improved later by the Belle \cite{Belle:2008jjb} and BABAR \cite{BaBar:2008wlm, BaBar:2010bul, BaBar:2012zfq} experiments who reported $\textrm{BR}(\tau^-\to\pi^-\eta\nu_\tau)<7.3\times 10^{-5}$ and $\textrm{BR}(\tau^-\to\pi^-\eta\nu_\tau)<9.9\times 10^{-5}$, respectively. An improvement can be established at Belle $\textrm{BR}(\tau^-\to\pi^-\eta\nu_\tau)<4.4\times 10^{-5}$ after analysing the full data set \cite{Ogawa:2020iwi}. 
 In the future, the Belle II experiment, which expects to produce a large data set containing $\sim 10^{10}$ tau pairs \cite{Belle-II:2018jsg}, can be able to measure for the first time the branching fraction of this decay channel. On the other hand, a stronger upper limit on the analogous $\tau^-\to \pi^-\eta'\nu_{\tau}$ decay has been reported by the BABAR Collaboration \cite{Workman:2022ynf}, namely B$(\tau^-\to \pi^-\eta'\nu_{\tau}) < 4.0 \times 10^{-6}$. To take advantage of these results in the search for NP, it is necessary that improved predictions of the branching fraction and other observables in $\tau^-\to \pi^-\eta\nu_{\tau}$ decay are obtained in the SM. This paper attempts to improve on this goal.
\section{The semileptonic  $\tau^- \to \pi^-\eta\nu_{\tau}$ amplitude }

It is well known that the semileptonic $\tau$ lepton decay into two pseudoscalar mesons is mediated by the vector current and described in terms of two form factors. For the $\tau^-(p_\tau)\to \pi^-(p_\pi)\eta(p_\eta)\nu_\tau(p_\nu)$ decay under consideration, the lowest order amplitude can be written in a factorizable form
\eq{
\mathcal{M}&=\frac{G_F V_{ud}}{\sqrt{2}}\, \ell^{\mu}\cdot {\cal H}_{\mu}, \label{amp-three}
}
where $\ell_{\mu}=\bar{u}(p_\nu)\gamma_\mu (1-\gamma_5)u(p_\tau)$ is the leptonic weak current and $V_{ud}$ is the element of the Cabibbo-Kobayashi-Maskawa matrix.  The hadronic matrix element ${\cal H}_{\mu}$ can be parametrized in terms of the  form factors $F_+^{\eta\pi}(s)$ and $F_0^{\eta\pi}(s)$, namely
\eq{ {\cal H}_{\mu} &=
\langle\eta(p_\eta)\pi^-(p_\pi)\vert \bar{d}\gamma_\mu u\vert 0\rangle \nonumber \\ &= -\sqrt{2}\bigg[\bigg(q'_\mu-\frac{\Delta_{\eta\pi}}{s}q_\mu\bigg)F_+^{\eta\pi}(s)+\frac{\Delta_{\eta\pi}}{s}q_\mu\, F_0^{\eta\pi}(s) \bigg] . \label{fpandf0}
}
In the above expressions we have defined $\Delta_{\eta\pi}=q\cdot q'=m_\eta^2-m_\pi^2$, as the product of the two independent momenta $q_\mu=(p_\eta+p_\pi)_\mu$ and $q'_\mu=(p_\eta-p_\pi)_\mu$. The form factors are Lorentz-invariant functions of $s=q^2$, the square of the invariant mass of the $\eta\pi$ system. The subindices $(+,0)$ in the form factors refer to the $L=1$ and $L=0$ angular momentum configurations of the hadronic pair, and they are called vector and scalar form factors, respectively.

The corresponding decay rate for this decay is the following
\eq{
\Gamma(\tau^-\to \pi^-\eta\nu_{\tau}) &= \frac{G_F^2 |V_{ud}|^2\, S_{EW}}{8(4\pi)^3m_{\tau}^3} \int^{m_{\tau}^2}_{(m_{\eta}+m_{\pi})^2}ds \frac{3\lambda^{1/2}(s,m_{\eta},m_{\pi})(m_{\tau}-s)^2}{s^3}\nonumber\\
& \ \ \ \times\bigg\{(2s+m_{\tau}^2)\lambda(s,m_{\eta}^2,m_{\pi}^2)|F^{\eta\pi}_{+}(s)|^2+3m_{\tau}^2\Delta_{\eta\pi}^2|F^{\eta\pi}_{0}(s)|^2\bigg\}\  \label{decayrate}
}
where $\lambda(x,y,z)=x^2+y^2+z^2-2(xy+xz+yz)$ and  $S_{EW}=1.0201$ is the universal short-distance electroweak correction \cite{Marciano:1988vm,Erler:2002mv}. Note that: 1) the vector and scalar form factors contributions do not interfere in the $\eta\pi^-$ mass distribution  \footnote{This is not true in the presence of photonic corrections because the boxes in loop corrections introduce a dependence of form factors upon an additional Mandelstam variable (see below).} and, 2) the contribution of the scalar form factor can be important due to the (large) mass splitting of $\pi^-$ and $\eta$ mesons.

In the limit that $G$-parity is an exact symmetry, the vector current cannot hadronize into the $\eta\pi^-$ state, thus $F_+^{\eta\pi}(s)=F_0^{\eta\pi}(s)=0$; consequently, this `second class' $\tau$ decay would be forbidden. As explained before, non-zero values of these form factors can be induced in the SM by isospin breaking (IB) effects, or by NP interactions, for instance, newly charged scalar or leptoquarks particles, etc. In the former case, they become suppressed since isospin breaking is expected to be at most a few percent compared to allowed modes ($\tau^- \to (\pi, 2\pi, 3\pi)^-\nu_{\tau}$). In the presence of NP, the amplitude can be suppressed by the scales associated with heavy mediators. Since SM and NP contributions may be suppressed at the same level, searching the $\tau^- \to \pi^-\eta\nu_{\tau}$ decay can be sensitive to the latter effects. Therefore, a good knowledge of the form factors is required in order to extract meaningful information on NP from future measurements of $\tau^-\to \pi^- \eta\nu_\tau$ observables.

In the SM, isospin symmetry is broken by both the mass difference of down and up quarks ($m_d-m_u$) and by the effects of electromagnetic (e.m.) interactions. Therefore, the induced form factors contain two terms (hereafter, we drop the superindex $\eta\pi$):
\eq{
F_{+,0}=F_{+,0}^{d-u}+F_{+,0}^{\ \rm e.m.} \label{totalFF}
}
\begin{table}[h!]
\begin{center}
\begin{tabular}{l|c|c|c}\hline
\multicolumn{1}{c}{Ref.} &  \multicolumn{1}{c}{$BR_S\times 10^5$} & \multicolumn{1}{c}{$BR_V\times 10^5$} &\multicolumn{1}{c}{$BR \times 10^5$}\\ \hline \hline
* (1982) Tisserant, Truong \cite{Tisserant:1982fc} & \multirow{2}{*}{1.60} & \multirow{2}{*}{0.26} &\multirow{2}{*}{1.86}\\
($\rho$, $a_0$ contributions) &  &  \\

\hline

* (1987) Bramon, Narison, Pich \cite{Pich:1987qq, Bramon:1987zb} & \multirow{2}{*}{1.50} & \multirow{2}{*}{0.12}&\multirow{2}{*}{1.62}\\
($\rho$, $a_0$ contributions) &  &  \\

\hline 

(1994) Neufeld, Rupertsberger \cite{Neufeld:1994eg}  & \multirow{2}{*}{1.06} & \multirow{2}{*}{0.15}&\multirow{2}{*}{1.21}\\
(NLO ChPT) &  &  \\

\hline 

*(2008) Nussinov, Soffer \cite{Nussinov:2008gx}  & \multirow{2}{*}{1.00} & \multirow{2}{*}{ 0.36 }&\multirow{2}{*}{1.36}\\
($\bar{q}q$ model)  &  &  \\

\hline

(2010) Paver, Riazuddin \cite{Paver:2010mz}  & \multirow{2}{*}{[0.2,2.3]} & \multirow{2}{*}{[0.2,0.6]}&\multirow{2}{*}{[0.4,2.9]}\\
($\rho$, $\rho'$, $a_0$, $a'_0$ VMD) &  &  \\

\hline

*(2012) Volkov, Kostunin \cite{Volkov:2012be}  & \multirow{2}{*}{$0.04$} & \multirow{2}{*}{ $ 0.44$}&\multirow{2}{*}{0.48}\\
(NJL model) &  &  \\

\hline

(2014) Descotes-Genon, Moussallam \cite{Descotes-Genon:2014tla}  & \multirow{2}{*}{ 0.20} & \multirow{2}{*}{0.13}&\multirow{2}{*}{0.33}\\
(ChPT + analyticity) &  &   \\

\hline

(2016) Escribano, Gonzalez, Roig \cite{Escribano:2016ntp}  & \multirow{2}{*}{ $1.41\pm 0.09$} & \multirow{2}{*}{ $0.26\pm 0.02$} & \multirow{2}{*}{$1.67\pm 0.09$ } \\
(RChT-3 coupled channels) &  & &\\

\hline \hline

\end{tabular}
\caption{Some of the previous estimates of the $\textrm{BR}(\tau^-\to\pi^-\eta\nu_\tau)$ reported in the literature that stem from isospin breaking in the $d-u$ quark mass difference. The subscript S (V) denotes the contribution of the scalar (vector) form factor to the total branching ratio (4th column). The spread of values  between predictions can be traced to the  different inputs and approximations among the various hadronization models. In addition, predictions marked with an asterisk use the narrow-width approximation for scalar and vector resonances. }\label{estimations}
\end{center} 
\end{table}
Most of the previous  works \cite{Leroy:1977pq, Tisserant:1982fc, Bramon:1987zb, Pich:1987qq, Diaz-Cruz:1991qbl, Bednyakov:1992af, Neufeld:1994eg, Nussinov:2008gx, Nussinov:2009sn, Paver:2010mz, Paver:2011md, Volkov:2012be, Descotes-Genon:2014tla, Escribano:2016ntp, Volkov:2021kus, Moussallam:2021flg} have focused on the calculation of the form factors induced mainly by the $m_d-m_u$ quark mass difference. 
The vector form factor is modelled in a way similar to the one of $\pi^-\pi^0$ channel, which is dominated by the $\rho(770)$ meson (including or not its excited states), followed by the $\pi^0\to \eta$ conversion due to $\pi^0\eta$ mixing \cite{Tisserant:1982fc, Bramon:1987zb, Pich:1987qq, Diaz-Cruz:1991qbl, Bednyakov:1992af, Neufeld:1994eg, Nussinov:2008gx, Nussinov:2009sn, Paver:2010mz, Paver:2011md, Volkov:2012be, Descotes-Genon:2014tla, Escribano:2016ntp, Volkov:2021kus, Moussallam:2021flg}.  On the other hand, the scalar form factor is assumed to be dominated by the scalar $a_0(980)$ meson \cite{Tisserant:1982fc, Bramon:1987zb, Pich:1987qq,  Neufeld:1994eg, Nussinov:2008gx,  Paver:2010mz, Volkov:2012be} or it can be calculated from the coupled channel rescattering $P_1P_2\to \eta\pi^-$ in the $J=0$ configuration \cite{Moussallam:2021flg, Escribano:2016ntp}. The results for the branching ratio that stem from the separation into vector and scalar terms, according to Eq. (\ref{decayrate}), are shown in Table \ref{estimations} as reported in the original references. The input data and approximations assumed in the different models are reflected in the spread of predictions for the branching fractions. This wide range of predictions needs to be tightened in order to draw a significant conclusion about NP from a future measurement.

As is well known, the isovector part of the electromagnetic quark current $j_{\mu}^{I=1}=(\bar{u}\gamma_{\mu}u-\bar{d}\gamma_{\mu}d)/2$, violates isospin (thus also $G$-) symmetry. To the best of our knowledge, IB effects induced by electromagnetic interactions have been considered only in Ref. \cite{Hernandez-Tome:2017pdc}, which turned out to be very small, of $O(\alpha^2)$ at the amplitude level.
In the next section, we present the IB effects induced at one-loop by virtual photons, which lead to an amplitude suppressed only at $O(\alpha)$. As already mentioned in the introduction, because the $\pi^0-\eta$ mixing\footnote{Strictly speaking, this $\pi^0-\eta$ mixing parameter  also contains a very small contribution from virtual photons through $\pi^0 \leftrightarrow (\rho\gamma, \omega\gamma)\leftrightarrow \eta$ loops, although they are different from the ones considered in this paper. } and the fine structure constant $\alpha$ turn out to be of the same order, one may expect {\it a priori} that both effects contribute to the amplitude with similar sizes. We attempt to test such a hypothesis in this paper.

\section{G-parity breaking induced by QED loops}

Here we focus on the computation of the IB amplitudes induced by virtual photons. For this purpose, we will use a resonance dominance model to describe the hadron and photon interaction vertices. This model has been used, for example, to compute the long-distance QED radiative corrections to $\tau^- \to (\pi, K)^-\nu_{\tau}$ decays in Ref. \cite{Decker:1994ea} or to study the observables of radiative $\tau^-\to \pi^-\pi^0\nu_{\tau}\gamma$ \cite{Flores-Tlalpa:2005msx} and $\tau^-\to \pi^- \eta\nu_{\tau}\gamma$ decays \cite{Guevara:2016trs}. Although this model does not satisfy the QCD behaviour of the form factors expected at short distances (a calculation which consistently implements this property in the calculation of QED radiative corrections to  $\tau^- \to (\pi, K)^-\nu_{\tau}$  was done in \cite{Arroyo-Urena:2021nil, Arroyo-Urena:2021dfe}), it captures the main features of photon-hadron interactions in the intermediate energy (resonance) region, which is relevant for the evaluation of loop-effects.

\begin{figure}[h!]
\begin{center}
\begin{tabular}{ccc}
\includegraphics[scale=.65]{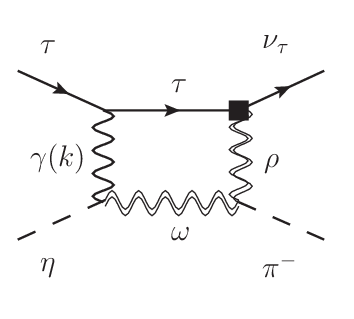}&
 \includegraphics[scale=.65]{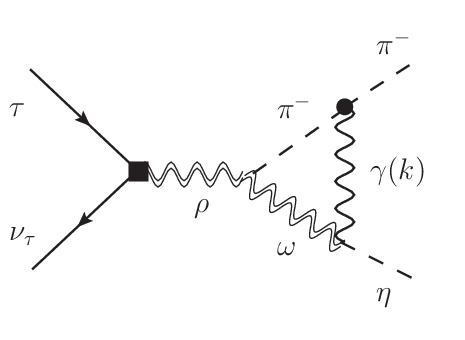} &  \includegraphics[scale=.65]{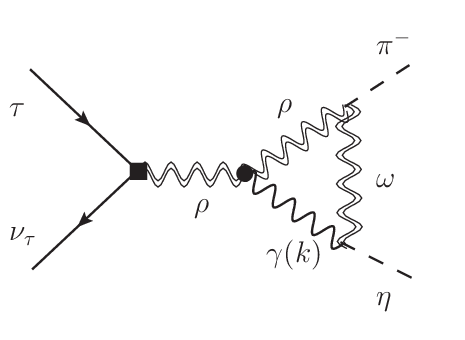} \\
{\small (a)} & {\small(b)} & {\small (c)}\\
 \end{tabular}
 \begin{tabular}{ccc}
\includegraphics[scale=.65]{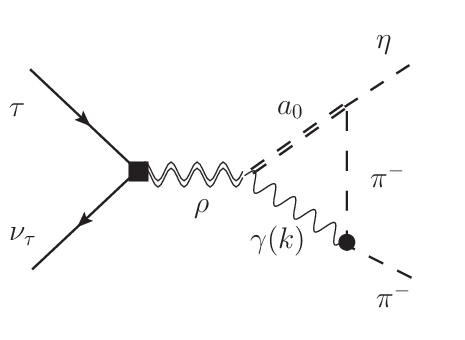}&
 \includegraphics[scale=.65]{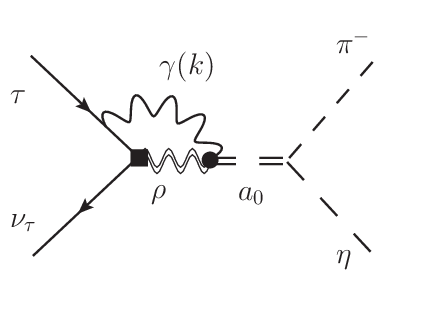} & \includegraphics[scale=.65]{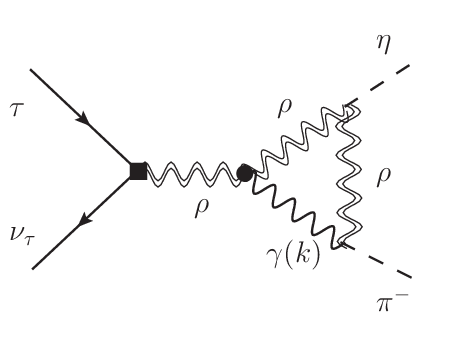}  \\
{\small (d)} & {\small(e)} & {\small (f)} 
 \end{tabular}
 \begin{tabular}{c}
      \includegraphics[scale=.65]{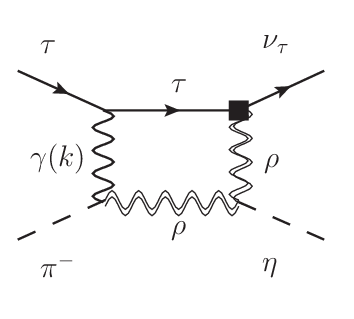}\\
      {\small (g)}
 \end{tabular}
 \caption{Feynman diagrams of the $\tau^-(p_\tau) \to \pi^-(p_\pi)\eta(p_\eta)\nu(p_\nu)$ decay induced by a virtual photon at one loop level. The black square stands for the integration out of the $W$ gauge boson, meanwhile, the circle represents the virtual photon interaction taking into account a squared momentum transfer dependence of the form given by Eq. (\ref{q2-dependence}).}\label{diagrams}
\end{center}
\end{figure}

At the leading order in photonic loops, the $\tau^-\to \pi^-\eta\nu_\tau$ decay can be induced in such a framework by the Feynman diagrams shown in Figure \ref{diagrams}. The presence of the virtual photon makes possible this decay at the one-loop level, in a similar way that the emission of a real photon in $\tau^- \to \pi^-\eta\nu_{\tau}\gamma$ avoids the $G$-parity constraint \cite{Guevara:2016trs}. Also, in this leading order, we include only the effects of the lowest lying vector $\left(\rho(770),\,\omega(782)\right)$ and scalar ($a_0(980)$) resonances. The effects of their excited states can play an important role above 1.4 GeV according to Ref. \cite{Paver:2010mz}, but we do not include them in this approximation given the lack of experimental information that would allow us to derive meaningful values of the relevant coupling.

\begin{table}[h!]
    \centering
    \begin{tabular}{c|c}\hline
       Coupling & Value \\
       \hline \hline 
       $g_\rho$ & $5.0\pm 0.1$\\
       $g_{\rho\omega\pi}$ &  $11.1\pm0.5$ GeV$^{-1}$\\       
       $e g_{\omega\eta\gamma}$ & $0.136\pm 0.016$ GeV$^{-1}$\\
       $e g_{\rho\pi\gamma}$ &  $0.219\pm 0.012$ GeV$^{-1}$\\
       $e g_{\rho a_0\gamma}$ & $0.092\pm 0.016$  GeV$^{-2}$\\
       $g_{\rho\rho\eta}$ & $7.9\pm 0.3$ GeV$^{-1}$\\
       $g_{\rho\rho\eta'}$ & $6.6\pm 0.2$ GeV$^{-1}$\\
       $eg_{\omega\eta'\gamma}$ & $0.13\pm 0.008$   GeV$^{-1}$\\
       $g_{a_0\pi\eta}$ & $2.2\pm 0.9$ GeV\\
       $g_{a_0\pi\eta'}$ & $\leq 0.22$ GeV\\
        \hline\hline
    \end{tabular}
    \caption{Strong and electromagnetic couplings of vector and scalar mesons used in our analysis (see reference \cite{Guevara:2016trs}).}
    \label{tab:couplings}
\end{table}

The evaluation of Feynman graphs shown in Figure \ref{diagrams} requires the effective vertices for $V_1V_2P$, $VP\gamma $, $VS\gamma$, and $SP_1P_2$ interactions, where $V,\ P$, and $S$ denote vector, pseudoscalar, and scalar mesons, respectively. As in Ref. \cite{Guevara:2016trs} (see also \cite{Bramon:1997va, Escribano:2020jdy, Titov:1999eu}), we use the following Feynman rules for the interaction of mesons and photons
\eq{
V_1^\mu(q_{1})\to V_2^\nu(q_{2})P(q_3)& :\quad i\,g_{{V_1}{V_2}P}\,\epsilon^{\mu\nu \alpha\beta}q_{2_\alpha}q_{3_\beta},\\
V^\mu(q_{1})\,\to \gamma^\nu(q_{2})P(q_3) &:\quad i\,g_{{V}\gamma P}\,\epsilon^{\mu\nu \alpha\beta}\,\,q_{2_\alpha}q_{3_\beta},\\
V^\mu(q_{1})\,\to \gamma^\nu(q_{2})S(q_3) &:\quad i\,g_{{V}\gamma S}\,\left(q_1\cdot q_2\, g^{\mu\nu}- q_2^{\mu}q_1^{\nu}\right),\\
S(q_1)\to P_1(q_2)P_2(q_3)& :\quad ig_{S{P_1}{P_2}}\ . \label{cou}
}
The $\rho^{-}-W$ coupling is defined as $\langle \rho^-(\epsilon^*)|\bar{d}\gamma_{\mu}u|0\rangle =(\sqrt{2}m_{\rho}^2/g_{\rho})\epsilon^{*}_{\mu}=f_{\rho}\epsilon^*_{\mu}$. The values for the above couplings required by our evaluation are obtained from other phenomenological analyses and are given in Table \ref{tab:couplings}. 

The electromagnetic vertex of the positively charged pion is defined as usual \cite{Bijnens:2002hp}
\eq{
\langle \pi^+(p_2)|J_{\mu}^{\rm em}(0)|\pi^+(p_1)\rangle &= e F_\pi^V(k^2) (p_1+p_2)_{\mu},\label{pi-pi-gamma}
}
where $J_{\mu}^{\rm em}(x)$ is the  electromagnetic current operator and $e$ the positron charge. The pion form factor  $F_\pi^V (k^2)$ is a function of the squared momentum transfer $k^2$ (where $k=p_2-p_1$) such that $F_\pi^V (0)=1$. 

Similarly, we will use the following expression for the electromagnetic matrix element of the  $\rho^+$ meson \cite{Nieves:1996ff, Hagiwara:1986vm}
\eq{
\langle \rho^+(p_2,\epsilon')|J_{\mu}^{\rm em}|\rho^+(p_1,\epsilon)\rangle &= e\,\epsilon'^{\beta *}\epsilon^{\alpha}\, \Gamma_{\alpha\beta\mu},\label{rho-rho-gamma}
}
where $\epsilon$ and $\epsilon'$ denote the initial and final polarization vectors, respectively.  The tensor $\Gamma^{\alpha\beta\mu}$ factor has the following Lorentz structure \cite{Lee:1962vm,LopezCastro:1999xg} 
\eq{
\Gamma^{\alpha\beta\mu}(k^2)&=(p_1+p_2)^\mu\, g^{\alpha\beta}\, \alpha(k^2)
+\left(g^{\mu\beta}\,k^\alpha-g^{\mu\alpha}\, k^{\beta}\right)\beta(k^2)\nonumber\\
&+(p_1+p_2)^\mu\, k^\alpha k^\beta \gamma(k^2)-p_{1}^{\alpha}g^{\mu\beta}-p_{2}^{\beta}\, g^{\mu\alpha}\ . \label{rho-rho-gamma}
}

The last two terms in the above equation do not contribute to on-shell vector mesons but are necessary to satisfy the Ward identities in the general case. The form factors $\alpha(k^2)$, $\beta(k^2)$, and $\gamma(k^2)$ are related to the static electromagnetic multipoles of the $\rho^+(770)$ vector meson   \cite{Brodsky:1992px}, respectively, as follows: $\alpha(0)=q=1$, $\beta(0)=\mu$ and $\gamma(0)=(1-\mu-\mathcal{Q})/2m_\rho^2$, where $q$ is the electric charge in units $e$, $\mu$ the magnetic dipole moment in units of $e/2 m_\rho$ and $\mathcal{Q}$ the electric quadrupole  in units of $e/m_\rho^2$. In this paper we will assume the canonical values \cite{GarciaGudino:2015ocw} $\alpha(0)=1, \beta(0)=2$ and $\gamma(0)=0$. We will comment later on the momentum transfer dependence of the form factors. 

It can be shown that, after some intermediate algebraic steps (see Appendix \ref{app:1}),  all the one-loop amplitudes corresponding to the diagrams in Figure \ref{diagrams} can be set into the following factorized generic form
\eq{
\mathcal{M}_{(i)}=\frac{G_F V_{ud}}{\sqrt{2}}C_{(i)}\, \ell_\mu \cdot \int \frac{d^dk}{(2\pi)^d} \frac{h^\mu_{(i)}}{D_{(i)}}, \label{generic-amplitude}
}
where the subindex $i=a,\,b,\, \cdots,\, g$ label the contribution of the diagrams in Figure \ref{diagrams}, and $C_{(i)}$ denote the product of couplings constants and (in some cases) meson propagators (see Appendix \ref{app:1}). 

It is interesting to note  that after the loop integration, the Lorentz structure of the amplitudes has an expression similar to Eq. (\ref{fpandf0}):
\eq{
\mathcal{M}_{(i)}&=\frac{G_F V_{ud}}{\sqrt{2}}\,\ell^\mu (-\sqrt{2})\bigg[\bigg(q'_\mu-\frac{\Delta_{\eta\pi}}{s}q_\mu\bigg)F_{+(i)}^{\textrm{\ e. m.}}(s, u ))+\frac{\Delta_{\eta\pi}}{s}q_\mu\, F_{0(i)}^{\ \rm e. m.}(s, u)) \bigg]\ .\label{amprad}
}
Note that the form factors generated by photonic loops are of order $\alpha$ and depend on an additional variable $u=(p_{\tau}-p_{\pi})^2$, where the latter originates from the box diagrams of Figure \ref{diagrams} \footnote{The explicit expressions for the $C_{(i)}$, $h^\mu_{(i)}$, and $D_{(i)}$ factors are reported in the Appendix \ref{app:1}. Similarly, the expressions for the $F_{\{+,0\}(i)}^{e. m.}$ factors in terms of the Passarino-Veltman functions are provided in Appendix \ref{app:2}.}. The form factors for the total amplitude induced by electromagnetic contributions are given by
 \eq{
 F_{\{+,0\}}^{\textrm{\ e. m.}}= \sum_{i=a}^{g} F_{\{+,0\}(i)}^{\textrm{\ e. m.}}.
 }
 Before presenting the numerical analysis, some relevant comments on our computation are in order. We have found that the triangular diagrams (b), (c), (e) and (f) in Figure \ref{diagrams}, have divergent behaviour in the limit where the photon-hadron vertices in Eqs. (\ref{pi-pi-gamma}) and (\ref{rho-rho-gamma}) are fixed at their zero momentum transfer values ($k^2=0$)\footnote{Diagram (d) is finite and well-behaved even in this approximation. However, we consider, for consistency, the $q^2$ dependence of the pion vector form factor in the evaluation of our estimation.}. In the vector meson dominance model considered in this paper, the interactions of the virtual photon with mesons are mediated by the exchange of vector mesons. Therefore, we will attach a factor ($m_\rho$ is the mass of the $\rho(770)$ vector meson)  
 \eq{
 {\cal F}_{i}(k^2)=\frac{m_\rho^2}{m_\rho^2-k^2}\label{q2-dependence}
 } 
to the electromagnetic vertices of charged ($\rho, \pi$) particles appearing in Figures \ref{diagrams}(b,c,f)  to describe their $k^2$ dependency. This factor is justified on the basis of many phenomenological descriptions of data and renders finite the divergent loop integrals (see for example \cite{Decker:1994ea}).

Similarly, for the diagram in Figure \ref{diagrams}(e), we assume that the virtual photon coupling in the $\rho^-a_0^-\gamma$ vertex occurs via the exchange of an $\omega(782)$ meson, the vector meson with suitable quantum numbers to couple to the $a_0^-\rho^-$ pair, which introduces an additional form factor ${\cal F}_{\rho a_0\gamma}(k^2)=m_\omega^2/(m_\omega^2-k^2)$ in the electromagnetic coupling.

\section{Electromagnetic contributions to the total rate.} 

In this section, we provide the results for the branching fraction of the $G$-parity breaking contribution to $\tau^- \to \pi^-\eta\nu_{\tau}$ that arise from the isospin breaking effects induced by the exchange of a virtual photon. We compare our results with the contributions due to the $m_d-m_u$ quark mass difference using the same model and approximations. For completeness, we also provide an estimate of these electromagnetic effects for the branching fraction of the analogous $\tau^- \to \pi^-\eta'\nu_{\tau}$ decay channel.

For later comparison, we first re-evaluate the branching fraction that stems from the $m_u-m_d$ quark mass difference through the $\pi^0\eta$ parameter $\epsilon_{\eta\pi}$. For this purpose, we also work in the framework of the meson resonance dominance model using the lowest-lying resonance states. Following Refs. \cite{Paver:2010mz, Paver:2011md}, the correctly normalized vector and scalar form factors that include the lowest lying and first excited resonances are the following:
\eq{
F_{+}^{u-d}(s)&=\epsilon_{\eta\pi}\times\frac{1}{1+\beta_\rho}\bigg[\left(\frac{m_{\rho}^2}{m_{\rho}^2-s-i m_{\rho}\Gamma_{\rho}(s)}+\frac{\beta_{\rho} m_{\rho'}^2}{m_{\rho'}^2-s-i m_{\rho'}\Gamma_{\rho'}(s)}\right)\bigg],\label{FpTree}\\
F_{0}^{u-d}(s)&=\epsilon_{\eta\pi}\times\frac{1}{1+\beta_{a_0}}\bigg[\left(\frac{m_{a_0}^2}{m_{a_0}^2-s-i m_{a_0}\Gamma_{a_0}(s)}+\frac{\beta_{a_0} m_{a'_0}^2}{m_{a'_0}^2-s-i m_{a'_0}\Gamma_{a'_0}(s)}\right)\bigg], \label{F0Tree}
}
where  $\beta_{\rho, a_0}$ are, in general,  complex parameters that describe the ratio of couplings of the excited/lightest mesons to the weak charged current and to the $\pi^-\pi^0$ meson pair. A similar expression, with $\epsilon_{\eta\pi}\to \epsilon_{\eta'\pi}$, holds for the form factors of $\tau^- \to \pi^-\eta'\nu_{\tau}$ decays.

In order to remain consistent with the approximation used in the loop calculations, we will use a single resonance to describe the form factors, namely we set $\beta_{\rho}=\beta_{a_0}=0$. Therefore, the only energy-dependent widths required in Eqs. (\ref{FpTree}, \ref{F0Tree}) are the following:
\eq{
\Gamma_{\rho}(s)&= \Gamma_{\rho}\left(\frac{m_{\rho}^2}{s}\right)^{5/2}\left(\frac{\lambda(s,m_{\pi}^2,m_{\pi}^2)}{\lambda(m_{\rho}^2,m_{\pi}^2,m_{\pi}^2)}\right)^{3/2}\theta(s-4m_{\pi}^2), \\
\Gamma_{a_0}(s)&=\Gamma_{a_0}\frac{m_{a_0}^2}{s}\frac{\lambda^{1/2}(s,m_{\eta}^2,m_{\pi}^2)}{\lambda^{1/2}(m_{a_0}^2,m_{\eta}^2,m_{\pi}^2)}\theta(s-(m_{\eta}+m_{\pi})^2),
} 
where $\theta(x)$ is the Heaviside functions and $\Gamma_{\rho, a_0}$ the on-shell widths. In our numerical evaluations, we use the masses and widths reported by the Particle Data Group \cite{Workman:2022ynf}, except for the scalar meson, where we assume $\Gamma_{a_0}=(75\pm 25)$ MeV to cover the range reported for this parameter in \cite{Workman:2022ynf}.  We also use the leading order expression for the isospin mixing parameter, namely $\epsilon_{\eta \pi} = (1.34)\times 10^{-2} $ \cite{Paver:2010mz}.

The values of the branching fractions obtained for the scalar and vector contributions owing to $m_u-m_d$ quark mass difference are reported in the line denoted as `$d-u$' in Table \ref{tab:ResultsBR}. The values in this Table for  \cite{Paver:2010mz}, differ slightly from the one reported in that reference because we use the correct phase-space for the energy-dependent width of the $\rho(770)\to \pi\pi$ decay.

\begin{table}[h!]
    \centering
    \begin{tabular}{c|c|c|c}\hline
         Diagram & $BR(\tau^-\to\pi^-\eta\nu_\tau)_S$ & $BR(\tau^-\to\pi^-\eta\nu_\tau)_V$ & $BR(\tau^-\to\pi^-\eta\nu_\tau)$ \\ \hline\hline
         (a) & $5.14\times 10^{-9}$& $6.14\times 10^{-9}$ & $9.21\, \times 10^{-9}$ \\
         (b) & $0$ & $3.87 \times 10^{-8}$& $3.87 \times 10^{-8}$\\     
         (c) & $1.00\times 10^{-8}$& $1.80\times 10^{-8}$ & $2.81\times 10^{-8}$\\
         (d) & $0$ & $6.05\times 10^{-10}$ & $6.05 \times 10^{-10}$\\
         (e) & $2.34\times 10^{-8}$& $0$ & $2.34\times 10^{-8}$\\
         (f) & $1.48\times 10^{-8}$&  $1.85\times 10^{-8}$& $3.33\times10^{-8}$ \\
         (g) & $5.81\times 10^{-9}$& $1.11\times 10^{-8}$& $1.46\times 10^{-8}$ \\ \hline
         e. m. & $1.64\times 10^{-7}$ & $4.17\times 10^{-8}$ & $2.15\times 10^{-7}$\\
          d-u &$1.48\times 10^{-5}$ & $3.21\times 10^{-6}$ & $1.80 \times 10^{-5}$\\
         d-u + e. m. &  $1.63\times 10^{-5}$&  $3.86\times 10^{-6}$&  $2.01\times 10^{-5}$\\ \hline\hline
    \end{tabular}
    \caption{Scalar (S) and vector (V) contributions to the branching ratio (BR) of  $\tau\to\pi^-\eta\nu_{\tau}$ from individual one-loop diagrams in Fig. \ref{diagrams}. The last three rows denote, the electromagnetic (e.m.), d-u quark mass difference contributions to the branching fraction and their sum (d-u+e. m.); respectively. 
    \label{tab:ResultsBR}}
\end{table}

The results shown in the upper part of Table \ref{tab:ResultsBR} correspond to the contributions of scalar (subindex $S$) and vector ($V$) form factors in   $BR(\tau^-\to\pi^-\eta\nu_{\tau})$, generated by the diagrams of Figure \ref{diagrams}. We note that the sum of scalar and vector contributions does not add up to the total branching ratio in the case of Figures \ref{diagrams} (a) and (g) because there is a small interference term between them that arise from the box diagrams (the induced form factors depend upon $(s, u)$ variables in this case). It is clear that the branching ratios of scalar and vector contributions induced by the pure photon loops are smaller by about two orders of magnitude with respect to the corresponding contributions induced by the $d-u$ quark mass difference. 

When we add the form factors generated by both sources of isospin breaking at the amplitude level according to Eq. (\ref{totalFF}), we get the branching ratios for $\tau^-\to \pi^-\eta\nu_{\tau}$ shown in the last row of Table \ref{tab:ResultsBR}. The shift produced by the photon corrections in the total rate becomes
\eq{
\frac{{|\rm BR_{d-u+e. m.}(\pi\eta)}-{\rm BR_{d-u}}(\pi\eta)|}{{\rm BR_{d-u}}(\pi\eta)}\approx 12\%\,  .
}
 Therefore, a measurement of the branching ratio of this decay at Belle II or at a $\tau$-charm factory with a $\sim$10\% uncertainty will require that all the effects of this order, in particular the ones due to the virtual photon, are explicitly taken into account in order to extract meaningful information on NP contributions.

\begin{table}[h!]
    \centering
    \begin{tabular}{c|c|c|c}\hline
         Diagram & $BR(\tau^-\to\pi^-\eta'\nu_\tau)_S$ & $BR(\tau^-\to\pi^-\eta'\nu_\tau)_V$ & $BR(\tau^-\to\pi^-\eta'\nu_\tau)$ \\ \hline\hline
         (a) & $8.72\times 10^{-10}$& $7.50\times 10^{-10}$ & $1.45\times 10^{-9}$ \\
         (b) & $0$ & $2.78\times 10^{-9}$& $2.78\times 10^{-9}$\\     
         (c) & $1.57\times 10^{-9}$& $2.16\times 10^{-9}$ & $3.75\times 10^{-9}$\\
         (d) & $0$ & $1.11\times 10^{-12}$ & $1.11\times 10^{-12}$\\
         (e) & $8.07\times 10^{-12}$& $0$ & $8.07\times 10^{-12}$\\
         (f) & $2.13\times 10^{-9}$& $1.35\times 10^{-9}$ &  $3.48\times 10^{-9}$\\
         (g) & $6.85\times 10^{-10}$& $1.68\times 10^{-9}$& $2.25\times 10^{-9}$ \\ \hline
         Total e. m. &  $1.73\times 10^{-8}$& $2.91\times 10^{-9}$& $2.08\times 10^{-8}$\\
         d-u &$5.76\times 10^{-8}$ & $2.16\times 10^{-9}$ & $5.98\times 10^{-8}$\\
         d-u + e. m. & $9.70\times 10^{-8}$& $8.65\times 10^{-9}$&  $1.06\times 10^{-7}$\\ \hline\hline
    \end{tabular}
    \caption{Same as Table \ref{tab:ResultsBR} but for the $\tau^-\to \pi^-\eta'\nu_\tau$ channel.}
    \label{tab:ResultsBR1}
\end{table}

Just for completeness, we also include the evaluation of the photon-loop contributions to the analogous $\tau^-\to\pi^-\eta'\nu_{\tau}$ decay. This decay is more suppressed than the $\eta\pi^-$ channel due to the smaller phase space available and also because the threshold for $\eta'\pi^-$ production is above the masses of light meson resonances. The relevant couplings entering the analogous diagrams in Figure \ref{diagrams} are shown in Table \ref{tab:couplings}. We use $\epsilon_{\eta'\pi}=(3\pm1)\times 10^{-3}$ \cite{Paver:2011md} for the $\pi^0-\eta'$ isospin mixing parameter. Our results\footnote{Here we take the same expressions given in Appendix \ref{app:2} by replacing the mass $m_{\eta}\to m_{\eta'}$ and the values for the relevant effective couplings.} are displayed in Table \ref{tab:ResultsBR1} following the same convention as in Table \ref{tab:ResultsBR}.

According to the results in Table \ref{tab:ResultsBR1}, in the $\eta'\pi^-$ channel the effects of the one-loop photon contributions are more important than in $\eta\pi^-$ relative to the one due to $m_d-m_u$.  When we add the effects of both sources of isospin breaking, the interference effects turn out to be larger than in the $\pi^-\eta$ case: 
\eq{
\frac{{|\rm BR_{d-u+e. m.}}(\eta'\pi)-{\rm BR_{d-u}}(\eta'\pi)|}{{\rm BR_{d-u}}(\eta'\pi)}\approx 78\%\,.
}
This result, however, should be taken with care because the exclusion of excited resonances involves two limitations: first, the $\pi^-\eta'$ system can be produced resonantly only with the inclusion of  higher resonances and, second, the current knowledge of the needed $\eta'$ couplings is still poor.

We end this section to comment on our approximations: 1) we have included only the lowest lying resonances in the calculation of the two sources of isospin breaking contributions; 2) we are taking isospin breaking in the $\pi^0-\eta-\eta'$ mixing parameters at the leading order. This allows us to keep the consistency of our approximations. The effects of excited resonances and next-to-leading order in mixing parameters can be important, as shown in Refs. \cite{Paver:2010mz, Paver:2011md}. More reliable information on the values of masses, widths and relevant branching ratios of excited resonances is necessary to account for these effects. We expect, however, that the relative size of form factors induced by electromagnetic interactions and u-d quark mass difference would not be largely affected.

\section{Conclusions}
The `second class' current $\tau^- \to \pi^-\eta\nu_{\tau}$ decay, forbidden in the limit of exact $G$-parity symmetry, can be a powerful tool to constrain/observe the effects of NP that generate effective scalar interactions at low energies \cite{Garces:2017jpz}. To achieve this goal, better estimates of the vector and scalar hadronic form factors induced by isospin breaking are needed. 

In this work we have evaluated for the first time the photon-loops contribution to this second-class decay, using a phenomenological resonance dominance model with the lowest-lying vector and scalar resonances. We find that those photon contributions can be as large as 12\% (78\% for the $\pi^-\eta'$ channel) of the total contribution. Thus, future measurements of the branching fraction of the $\pi^-\eta$ channel within a $\sim$10\% error would require the inclusion of the photon-loop contributions calculated in this paper in order to draw meaningful conclusions on possible NP contributions.

Our calculation can be improved by including the effects of the excited resonances. Currently, however, the lack of reliable information on some of the relevant couplings needed for loop calculations prevents us to include them in our calculations. 

\section*{Acknowledgements}

The work of G.H.T. is funded by \emph{Estancias Posdoctorales por México para la Formación y Consolidación de las y los Investigadores por México, Conahcyt}. G.L.C. acknowledges the financial support of Conahcyt through Ciencia de Frontera project CF-2019-428218. D.P.S. is grateful to Conahcyt for  support through a doctoral fellowship.

\appendix
\section{Calculation of the form factors $F_\pm^{\rm e.m.}$ induced by a photon-loop}\label{app:1}

In this Appendix we relate the form factors that describe the hadronic matrix elements of $\tau^-(p_\tau)\to \eta(p_\eta)\pi^-(p_\pi)\nu_{\tau}(p_{\nu})$ decay. 
As it will be shown below, the amplitudes induced by photon loops can be written in a factorized form similar to Eq. (\ref{amp-three}). We find it convenient to introduce first a simpler parametrization of the hadronic matrix element as follows 
\eq{
{\cal H}_{\mu}^{\rm e.m.} = -\sqrt{2}
\left\{ F_+^{\rm e.m.} (s,u) q'_{\mu}+F_-^{\rm e.m.} (s,u) q_{\mu} \right\} \label{basispm}
} 
where $q'=p_\eta-p_\pi, \  q=p_\eta+p_\pi$. Given the contribution of box diagrams, the form factors acquire a dependence upon the variable $u=(p_\tau-p_\pi)^2$. This set of form factors is related to the ones used in Eq. (\ref{fpandf0}) by means of 
\eq{
F_{0}^{\textrm{e.m.}}=F_{+}^{\textrm{e.m.}}+\frac{s}{\Delta_{\eta\pi}} F_{-}^{\textrm{e.m.}} \ . \label{ffminusto0}
}
In this appendix, we evaluate the form factors in the basis provided by Eq. (\ref{basispm}) and then compute the scalar form factor using Eq. (\ref{ffminusto0}). 

\subsection{Contribution of diagrams (a), (e) and (g)}

The amplitudes for these diagrams ($i=a, e, g$) in Figure \ref{diagrams} have the general form
\eq{
\mathcal{M}_{(i)}= \frac{G_F V_{ud}}{\sqrt{2}} C_{(i)}\int \frac{d^dk}{(2\pi)^d} \frac{\ell_{\mu\nu}\cdot h^{\mu\nu}_{(i)}}{\mathcal{D}_{(i)}},
\label{amp-int}
}
where $\ell_{\mu\nu}=\bar{u}(p_\nu)\gamma_\mu \gamma_(1-\gamma_5)[(\cancel{p_\tau}+\cancel{k})+m_\tau]\gamma_\nu u(p_\tau)$ is the leptonic tensor, and the $O(\alpha)$ coefficients $C_{i}$ are the product of coupling constants and resonance propagators (see Appendix \ref{app:amplitudes}). The hadronic tensors  $h^{\mu\nu}_{(i)}$ have the  following forms (see the definitions of the four-rank tensors $T$ and $\hat{T}$ in Appendix \ref{app:amplitudes})
\eq{
h^{\mu\nu}_{(a)}& = \epsilon^{\mu}\,{}_{\mu_1\mu_2\mu_3}\,\epsilon^{\mu_1\nu}{}_{\mu_4\mu_5}\,T^{\mu_2\mu_4\mu_3\mu_5},\\
h^{\mu\nu}_{(e)}& =\left[k\cdot (q+k)g^{\mu\nu}-k^{\mu}(k+q)^\nu\right],\\
h^{\mu\nu}_{(g)}& = \epsilon^{\mu}\,{}_{\mu_1\mu_2\mu_3}\,\epsilon^{\mu_1\nu}{}_{\mu_4\mu_5}\,\hat{T}^{\mu_2\mu_4\mu_3\mu_5}.
}
Using the Dirac equation and the Chisholm identity we have the following identity \footnote{The Chisholm identity used here reads $\gamma_\mu\gamma_\lambda\gamma_\nu (1-\gamma_5)=\alpha_{\mu\nu\lambda\sigma}\gamma^{\sigma}(1-\gamma_5)$}..
\eq{
\ell_{\mu\nu}=\ell^\sigma\left[2g_{\mu\sigma}(q+p_{\nu})_{\nu}+\alpha_{\mu\nu\lambda\sigma}k^\lambda \right],
}where $\alpha_{\mu\nu\lambda\sigma}\equiv g_{\mu \lambda}g_{\nu \sigma}+ g_{\lambda \nu}g_{\mu \sigma}-g_{\mu \nu}g_{\lambda \sigma}+i\epsilon_{\mu\nu\lambda\sigma}$. The integral in Eq. (\ref{amp-int}) can  be set as
\eq{
\int \frac{d^dk}{(2\pi)^d} \frac{\ell_{\mu\nu}\cdot h^{\mu\nu}_{(i)}}{\mathcal{D}_{(i)}}=& \ell_{\sigma} \int \frac{d^dk}{(2\pi)^d} \frac{h^{\sigma}_{(i)}}{\mathcal{D}_{(i)}} ,\nonumber\\
=&\ell_\sigma\left[f_{+(i)}^{\eta\pi}\, q'^\sigma + f_{-(i)}^{\eta\pi}\, q^\sigma +f_{(i)}^{\nu_\tau}\, p_{\nu_\tau}^\sigma + i f_{(i)}^{\epsilon}\epsilon_{\mu\nu\lambda\sigma}q'^\mu q^\nu p_{\nu_\tau}^\lambda\right].\label{inthmunu}
}
Notice that the third term in the above expression vanishes owing to $\ell_\sigma p_{\nu_\tau}^\sigma=0$. Moreover, the last term can be rewritten as follows
\eq{
if_{(i)}^{\epsilon}\ell^\sigma\cdot \epsilon_{\mu\nu\lambda\sigma}q'^\mu q^\nu p_{\nu_\tau}^\lambda=f_{(i)}^{\epsilon} \ell_{\sigma}\left[(q' \cdot p_\nu) q^\sigma - (q \cdot p_\nu) q'^\sigma  \right].
}
Therefore, the contribution  of diagrams (a), (e), and (g) in Figure \ref{diagrams} to the form factors $F_\pm^{\rm e.m.}(s,u)$ are given by
\eq{
F_{+(i)}^{\textrm{e.m}}=-\frac{C_{(i)}}{16\pi^2\sqrt{2}}\bigg[f_{+(i)}^{\eta\pi}+f_{(i)}^{\epsilon}(q \cdot p_\nu) \bigg], \quad F_{-(i)}^{\textrm{e.m}}=-\frac{C_{(i)}}{16\pi^2\sqrt{2}}\bigg[f_{-(i)}^{\eta\pi}-f_{(i)}^\epsilon (q' \cdot p_\nu) \bigg].
} 

\subsection{Contribution of diagrams (b), (c), (d), (f)}
As it can be seen from a direct inspection, the amplitudes for diagrams in Figures \ref{diagrams}(b, c, d, f) can be factorized as in eq. (\ref{amp-three}). The hadronic matrix elements ${\cal H}^{\rm e.m}_{(i) \mu}$, in this case, are proportional to the loop integrals in Eq. (\ref{generic-amplitude}), namely (for $i=b,\ c,\ d,\ f$)
\eq{
\int \frac{d^dk}{(2\pi)^d} \frac{h^\mu_{(i)}}{\mathcal{D}_{(i)}}=f_{+(i)}^{\eta\pi}\, q'^\mu+ f_{-(i)}^{\eta\pi}\, q^\mu\, ,\label{inthmu}
}where the factors $f_{\pm(i)}^{\eta\pi}$ are given in terms of Passarino-Veltman functions (see below). Then it is immediate to identify that
\eq{
F_{\pm (i)}^{\textrm{e.m}}=-\frac{C_{(i)}}{16 \pi^2\sqrt{2} }f_{\pm (i)}^{\eta\pi}.
}

\section{One loop amplitudes}\label{app:amplitudes}

In this appendix, we report the expressions of the factors $h_{(i)}$, $D_{(i)}$, and $C_{(i)}$ in Eq. (\ref{generic-amplitude}) that appear in the amplitudes for the different diagrams in Fig.  \ref{diagrams}. First, the hadronic  $h_{(i)}^\mu$ term in the integrand of Eq. (\ref{generic-amplitude}) are given as follows
\eq{
h_{(a)}^\mu &=\bigg[2g^{\mu\,{\mu_1}}(q+p_\nu)^{\mu_2}+\alpha^{\mu_1\mu_2\mu_3\mu}\,k_{\mu_3}\bigg]\epsilon_{\mu_1\mu_4\mu_5\mu_6}\,\epsilon^{\mu_4}\,_{\mu_2\mu_7\mu_8}\,T^{\mu_5\mu_7\mu_6\mu_8}, \\
h^\mu_{(b)} & = 2\epsilon^{\mu \mu_1\mu_2\mu_3}\,\epsilon_{\mu_1\mu_4\mu_5\mu_6}\,q_{\mu_2}\, \left(k-\frac{q+q^{\prime}}{2}\right)_{\mu_3}\, \left(\frac{q-q'}{2}\right)^{\mu_4}\,\left(\frac{q+q'}{2}\right)^{\mu_5}\,k^{\mu_6}\, ,\label{hb}\\
h^{\mu}_{(c)}& = \left[g_{\mu\mu_1}-\frac{q_{\mu}\,q_{\mu_1}}{m_{\rho}^2}\left(1+\frac{i\,m_{\rho}\Gamma_{\rho}(s)}{s}\right) \right]\,\Gamma^{\mu_1\mu_3\mu_2}(0)\, \epsilon_{\mu_3\mu_4\mu_5\mu_6}\,\epsilon^{\mu_4}{}_{\mu_2\mu_7\mu_8}\, T^{\mu_5\mu_7\mu_6\mu_8},\\
h_{(d)}^\mu &=\bigg[(k.q)g^{{\mu}\,{\mu_2}}-k^{\mu}q^{\mu_2}\bigg](k+q-q')_{\mu_2}, \\
h_{(e)}^\mu &=\bigg[2g^{\mu\,{\mu_1}}(q+p_\nu)_{\mu_2}+\alpha^{{\mu_1}{\mu_2}{\mu_3}\mu}\,k^{\mu_3}\bigg]\bigg[k.(q+k)\, g_{{\mu_1}{\mu_2}}-k_{\mu_1}(q+k)_{\mu_2}\bigg],\\
h_{(f)}^\mu &= \left[g_{\mu\mu_1}-\frac{q_{\mu}\,q_{\mu_1}}{m_{\rho}^2} \left(1+\frac{i\,m_{\rho}\Gamma_{\rho}(s)}{s}\right)\right]\, \Gamma^{\mu_1\mu_3\mu_2}(0)\, \epsilon_{\mu_3\mu_4\mu_5\mu_6}\,\epsilon^{\mu_4}{}_{\mu_2\mu_7\mu_8}\, \hat{T}^{\mu_5\mu_7\mu_6\mu_8},\\
h_{(g)}^{\mu}&=\bigg[2g^{\mu\,{\mu_1}}(q+p_\nu)^{\mu_2}+\alpha^{\mu_1\mu_2\mu_3\mu}\,k_{\mu_3}\bigg]\epsilon_{\mu_1\mu_4\mu_5\mu_6}\,\epsilon^{\mu_4}\,_{\mu_2\mu_7\mu_8}\,\hat{T}^{\mu_5\mu_7\mu_6\mu_8}, 
}
where we have defined the four-rank tensors
\eq{T^{\mu_5\mu_7\mu_6\mu_8}&=\left(\frac{q-q^{\prime}}{2}\right)^{\mu_5}\left(\frac{q+q^{\prime}}{2}\right)^{\mu_7}\left(k+\frac{q+q^{\prime}}{2}\right)^{\mu_6} k^{\mu_8} \\
\hat{T}^{\mu_5\mu_7\mu_6\mu_8}&=\left(\frac{q+q^{\prime}}{2}\right)^{\mu_5}\left(\frac{q-q^{\prime}}{2}\right)^{\mu_7}\left(k+\frac{q-q^{\prime}}{2}\right)^{\mu_6} k^{\mu_8}\ .
}
The denominators that appear in the integrand of Eq. (\ref{generic-amplitude}) are the following
\eq{
\mathcal{D}_{(a)}&=k^2\,G(k+q+p_{\nu},m_{\tau})G(k+q,m_{\rho})G\left(k+\frac{q+q'}{2},m_{\omega}\right)\,,\\
\mathcal{D}_{(b)}&=k^2\,G(k,m_{\rho})G\left(k+\frac{q-q'}{2},m_{\pi}\right)G\left(k-\frac{q+q'}{2},m_{\omega}\right)\,,\\
\mathcal{D}_{(c)}&=k^2\,G(k,m_{\rho})G(k+q,m_{\rho})G\left(k+\frac{q+q'}{2},m_{\omega}\right)\,,\\
\mathcal{D}_{(d)}&=k^2\,G(k,m_{\rho})G(k+q,m_{a_0})G\left(k+\frac{q-q'}{2},m_{\pi}\right)\,,\\
\mathcal{D}_{(e)}&=k^2\,G(k,m_{\omega})G(q+k,m_{\rho})G(k+q+p_{\nu},m_{\tau})\,,\\
\mathcal{D}_{(f)}&=k^2\,G(k,m_{\rho})\,G(k+q,m_{\rho})G\left(k+\frac{q-q'}{2},m_{\rho}\right)\,,\\
\mathcal{D}_{(g)}&=k^2\,G(k+q+p_{\nu},m_{\tau})G(k+q,m_{\rho})G\left(k+\frac{q-q'}{2},m_{\rho}\right)\,,
}where $G(k,m)\equiv k^2-m^2.$
Finally, the $C_{(i)}$ coefficients in Eq. (\ref{generic-amplitude}) are of $O(e^2)$ as expected and are given by
\eq{
C_{(a)}&=-e\,f_{\rho}\,g_{\rho\omega\pi}\,g_{\omega\gamma\eta},\\
C_{(b)}&=\frac{m_\rho^2 \,e \,f_{\rho}\,g_{\rho\omega\pi}\,g_{\omega\gamma\eta}}{s-m_\rho^2+im_{\rho}\Gamma_{\rho}(s)},\\
C_{(c)}&=-C_{(b)},\\
C_{(d)}&=-\frac{m_\rho^2 e f_\rho\,g_{\rho a_0 \gamma}\, g_{a_0\eta\pi}}{s-m_\rho^2+im_{\rho}\Gamma_{\rho}(s)},\\
C_{(e)}&=-\frac{m_\omega^2\, e f_\rho\,g_{\rho a_0 \gamma}\, g_{a_0\eta\pi}}{ s-m_{a_0}^2+im_{a_0}\Gamma_{a_0}(s)}\, ,\\
C_{(f)}&=-\frac{m_\rho^2\, e\,f_{\rho}\,g_{\rho\rho\eta}\,g_{\rho\gamma\pi}}{s-m_\rho^2+im_{\rho}\Gamma_{\rho}(s)}\, ,\\
C_{(g)}&=-e\,f_{\rho}\, g_{\rho\rho\eta}\, g_{\rho\gamma\pi}, 
}
where $f_\rho$ is defined one line below Eq. (\ref{cou}).  

\section{Loop Functions}
\label{app:2}

We have used \emph{Package-X} \cite{Patel:2015tea} to express our results. The definition and decomposition of the Passarino-Veltman functions reported here can be found in Appendix A of reference \cite{Arroyo-Urena:2021dfe}. Our results are reported as follows

Diagram (a):
\eq{
f_{+(a)}^{\eta\pi}&=-\frac{1}{4} \bigg[s (5 D_{001} + 8 D_{00}) + (\Delta_{\eta\pi}^2 - \xi s) (D_{112} + D_{113} + D_{122} + 2 D_{123} + D_{12} \nonumber\\
&+ D_{133} + D_{13}) - 3 \xi D_{001} + \chi (4 D_{00} - \xi (D_{113} + 2 D_{123} + 2 D_{133} + D_{13}) + \chi \nonumber\\
&\times(2 D_{23} - D_{133} - D_{33})) - \chi' (4 D_{00} - 4 D_{003} + \chi (2 D_{23} + D_{33})) + \chi'^2 D_{133}  \nonumber\\
&+ \Delta_{\eta\pi} (2 D_{001} + 4 D_{002} + 4 D_{003} + \chi' (D_{113} + 2 (D_{123} + D_{133}) + D_{13})) + (\Delta_{\eta\pi}  \nonumber\\
&\times \chi- \chi' s) (2 D_{12} - D_{113} - D_{13} + 2 D_{2} + 2 D_{22} + 3 D_{23} + D_{33}) \bigg],\\
f_{-(a)}^{\eta\pi}&=-\frac{1}{4}\bigg[(3 s - 5 \xi) (D_{001} + 2 (D_{002} + D_{003})) + (\Delta_{\eta\pi}^2 - \xi s) (D_{112} + D_{113} + 3 D_{122} \nonumber\\
&+ 6 D_{123} + D_{12} + 3 D_{133} + D_{13} + 2 (D_{222} + 3 D_{223} + D_{22} + 3 D_{233} + 2 (D_{23} \nonumber\\
&+ D_{33}))) - \chi (4 D_{003} - 4 D_{00} + \chi (D_{133} + 2 (D_{233} + D_{33}))) - 2 \Delta_{\eta\pi} (D_{001} \nonumber\\
&+ 4 D_{00})+ \chi' (-4 D_{00} + \chi (D_{33} - 2 D_{23})) + (\chi' \Delta_{\eta\pi} - \chi \xi) (D_{113} + 4 D_{123} + 2 D_{12} \nonumber\\
&+ 4 D_{133} + D_{13} + 4 D_{223} + 2 D_{22} + 8 D_{233} + 5 D_{23} + 2 D_{2} + 7 D_{33}) + (s \chi' \nonumber\\
& - \Delta_{\eta\pi} \chi) (D_{113}+ 2 D_{123} + 2 D_{133} + D_{13} + 2 D_{23} + 2 D_{33}) + \chi'^2 (D_{133} + 2 (D_{233} \nonumber\\
&+ D_{23})+ 3 D_{33})\bigg],\\
f_{(a)}^{\epsilon}&=-\frac{-1}{8} \bigg[24 D_{003} + s (4 D_{133} + D_{13} + 4 D_{223} + 8 D_{233} + 2 D_{23} + 6 D_{33}) +  2 \Delta_{\eta\pi} \nonumber\\
&\times(2 D_{133} + D_{13} + D_{23} + D_{33}) + 4 (s + \Delta_{\eta\pi})  D_{123} + \xi D_{13} + (s + 2 \Delta_{\eta\pi} + \xi) \nonumber\\
&\times D_{113} +  2 \chi (2 D_{133} + 4 D_{233} + 5 D_{33}) + 2 \chi' (2 D_{133} + D_{33})\bigg].}
Diagram (b):
\eq{
f_{+(b)}^{\eta\pi}&=-2m_{\rho}^2\,s\,\widetilde{D}_{00} ,\label{fpb}\\
f_{-(b)}^{\eta\pi}&=2m_{\rho}\Delta_{\eta\pi}\widetilde{D}_{00}.\label{fmb}
}
Diagram (c):
\eq{
f_{+(c)}&=\frac{m_{\rho}^2}{2}\bigg[(\xi s-\Delta^2) (\widehat{D}_{12}+ \widehat{D}_{122}+ \widehat{D}_{112})-2 (\Delta+3 s) \widehat{D}_{00}-4 \Delta \widehat{D}_{002}\nonumber\\
&+(-2 \Delta+3 \xi-5 s) \widehat{D}_{001}\bigg],\\
f_{-(c)}&=\frac{1}{2}\bigg[ (m_{\rho}^2 (6 \Delta+5 \xi-3 s)+\alpha (2 \Delta^2+3 s^2-5 \xi s)) \widehat{D}_{00}+2 (m_{\rho}^2 (5 \xi-3 s)\nonumber\\
&+\alpha (2 \Delta^2+3 s^2-5 \xi s)) \widehat{D}_{002}+(m_{\rho}^2 (2 \Delta+5 \xi-3 s)+\alpha (2 \Delta^2-3 \Delta \xi+3 s^2\nonumber\\
&+3 \Delta s-5 \xi s)) \widehat{D}_{001}+ (m_{\rho}^2-\alpha s) (\xi s-\Delta^2)( 3 \widehat{D}_{22}+2 \widehat{D}_{222}+ \widehat{D}_{2})+(\xi s-\Delta^2)\nonumber\\
&\times((2 m_{\rho}^2-\alpha (\Delta+2 s)) \widehat{D}_{12}+ (3 m_{\rho}^2-\alpha (\Delta+3 s)) \widehat{D}_{122}+ (m_{\rho}^2-\alpha (\Delta+s)) \widehat{D}_{112})\bigg].
}
Diagram (d):
\eq{
f_{+(d)}^{\eta\pi}&=m_{\rho}^2 s \,D_1(s,m_{\eta}^2,m_{\pi}^2,0;m_{\pi}^2,s;0,m_{a_0},m_{\pi},m_{\rho}),\nonumber\\
f_{-(d)}&=-m_{\rho}^2\Delta_{\eta\pi}\,D_1(s,m_{\eta}^2,m_{\pi}^2,0;m_{\pi}^2,s;0,m_{a_0},m_{\pi},m_{\rho}).
}
Diagram (e):
\eq{
f_{+(e)}&=0,\nonumber\\
f_{-(e)}&=m_{\omega}^2 \bigg[-3 m_{\tau}^2 \overline{D}_{222}-(2 m_{\tau}^2+s) (\overline{D}_{22}-3  \overline{D}_{12}-3  \overline{D}_{122})-(m_{\tau}^2+2 s)\nonumber\\
&\times(\overline{D}_{11}-3  \overline{D}_{112})-3 s \overline{D}_{111}-18 (\overline{D}_{002}+ \overline{D}_{001})+(s-m_{\tau}^2) \overline{D}_{1}\bigg].
}
Diagram (f):
\eq{f_{+(f)}^{\eta\pi}&=-f_{+(c)}^{\eta\pi}(m_{\pi}\leftrightarrow m_{\eta},m_{\omega}\rightarrow m_{\rho}),\nonumber\\
f_{-(f)}^{\eta\pi}&=f_{-(c)}^{\eta\pi}(m_{\pi}\leftrightarrow m_{\eta},m_{\omega}\rightarrow m_{\rho}).}
Diagram (g):
\eq{f_{+(g)}^{\eta\pi}&=-f_{+(a)}^{\eta\pi}(m_{\pi}\leftrightarrow m_{\eta},m_{\omega}\rightarrow m_{\rho},u\rightarrow t),\nonumber\\
f_{-(g)}^{\eta\pi}&=f_{-(a)}^{\eta\pi}(m_{\pi}\leftrightarrow m_{\eta},m_{\omega}\rightarrow m_{\rho},u\rightarrow t),\nonumber\\
f_{-(g)}^{\epsilon}&=-f_{-(a)}^{\epsilon}(m_{\pi}\leftrightarrow m_{\eta},m_{\omega}\rightarrow m_{\rho},u\rightarrow t).}
In the above expressions we have defined $t\equiv (p_{\tau}-p_{\eta})^2=m_{\eta}^2+m_{\pi}^2+m_{\tau}^2-s-u$, $\xi \equiv (q^{\prime})^2=2(m_{\eta}^2+m_{\pi}^2-s)$, $\chi \equiv p_{\nu} \cdot q =(m_{\tau}^2-s)/2$, $\chi^{\prime} \equiv p_{\nu} \cdot q^{\prime}=(2m_{\pi}^2+m_{\tau}^2-s-2u)/2$ and $\alpha\equiv 1+ i m_{\rho} \Gamma_{\rho}/s$. Moreover, we use the following notation to define the arguments of the Passarino-Veltman functions 
\eq{
D_i&\equiv D_i(m_{\eta}^2,m_{\pi}^2,0,m_{\tau}^2;s,u;0,m_{\omega},m_{\rho},m_{\tau}),\\
\widetilde{D}_{i}&\equiv \widetilde{D}_i(m_{\pi}^2,s,m_{\eta}^2,0;m_{\eta}^2,m_{\pi}^2;0,m_{\pi},m_{\omega},m_{\rho}),\\
\widehat{D}_{i}&\equiv \widehat{D}_i (m_{\eta}^2,m_{\pi}^2,s,0;s,m_{\eta}^2;0,m_{\omega},m_{\rho},m_{\rho}),\\
\overline{D}_i&\equiv \overline{D}_i(s,0,m_{\tau}^2;m_{\tau}^2,s;0,m_{\rho},m_{\tau},m_{\omega}).
}

\bibliography{biblio}{}
\bibliographystyle{unsrt}

\end{document}